\documentclass[10pt,conference,letterpaper]{IEEEtran}
\usepackage[T1]{fontenc}

\IEEEoverridecommandlockouts
\usepackage{cite}
\usepackage{amsmath,amssymb,amsfonts}
\usepackage{textcomp}
\usepackage{xcolor}
\usepackage{graphicx}
\usepackage{algorithm}
\usepackage{algpseudocode}
\usepackage[capitalise]{cleveref}
\usepackage{booktabs}
\usepackage{microtype}
\usepackage{multirow}
\usepackage{subcaption}
\usepackage[normalem]{ulem}
\begin{document}

\title{Denial of Service Attacks on \\Beam Allocation
in Large LEO Constellations
\thanks{This version of the paper was submitted to IEEE INFOCOM 2027 on \\ July 31, 2026.}
\thanks{\textsuperscript{\dag}Corresponding author: Ali Abbasi (abbasia@usc.edu).}}

\author{\IEEEauthorblockN{Ali Abbasi\textsuperscript{*,\dag}, Pasha Barahimi\textsuperscript{*}, Marco Paolieri\textsuperscript{*} and Leana Golubchik\textsuperscript{*}}
\IEEEauthorblockA{\textsuperscript{*}\textit{University of Southern California}, Los Angeles, USA \\
\{abbasia, barahimi, paolieri, leana\}@usc.edu}
}
\maketitle

\begin{abstract}
We study the vulnerability of beam allocation approaches in large low-Earth-orbit (LEO) satellite constellations to adversarial demand inflation resulting in denial-of-service attacks. We consider an attack model in which an adversary places (injects) additional terminals and evaluate these attacks against greedy and max-min-fair beam allocation algorithms, each combined with either equal or demand-proportional intra-cell capacity sharing. Vulnerability is measured using mean demand satisfaction, coverage, and the fraction of legitimate users receiving at least a threshold fraction of their requested capacity under several attack heuristics. There is a trade-off for the system between robustness and performance, meaning the algorithm with the most vulnerability to the attacks (greedy-proportional) has the highest performance before the attacks. Demand-proportional sharing increases the fraction of under-served users up to 34\% under both allocation algorithms, whereas equal sharing limits the increase up to 13\%. The sharing rule also changes the most damaging attack allocation in nearly all cases: demand-proportional sharing rewards concentrating injected terminals in the most loaded cells, while equal sharing performs better when distributing these terminals across the least loaded cells. Furthermore, the max-min-fair beam allocation algorithm forces the attacker to use a higher number of targeted cells to result in extensive damage to the system.
\end{abstract}


\section{Introduction}
Low-Earth-orbit (LEO) constellations are expanding rapidly, with several large constellations consisting of thousands of
satellites already deployed and providing commercial broadband service nowadays.
Measurement studies in numerous countries report latency and throughput performance comparable to terrestrial cellular networks
\cite{mohan2024starlink,izhikevich2024democratizing}. Such success has changed what these networks were meant to do previously. In recent work, they are being evaluated as national
failover capacity for submarine cable outages, and they are assumed to be the only connectivity option for remote regions \cite{bhosale2025leo}. As LEO networks
become widely used, their security and resilience to adversarial
attacks are increasing concerns.

A LEO operator cannot always add capacity when demand
increases, as downlink capacity is produced by a fixed number of phased-array
beams per satellite, drawn from a small pool of orthogonal channels that
must be reused across geographically separated cells 
\cite{zhang2023syslevel,lin2023multi}. This will cause the provided capacity to the user to be limited to available beams from visible satellites, which are dynamic over time, unlike a terrestrial ISP. A satellite crossing a
region can illuminate only a small number of cells at a time and must
choose which cells to serve. The choice is made by a beam allocation
algorithm which considers the aggregate demand of each cell. Capacity is therefore not only finite,
but also affected by demand. Whoever influences the demand
can influence where the constellation points its beams, how the resulting
capacity is divided among terminals underneath, and therefore how satisfied the users will be. In this paper we study what an adversary can do when demand is its attack surface.

Security work on LEO networks has focused on different approaches. For the
physical layer, there are studies on jamming, spoofing, and
interception of the radio links \cite{yue2023leosec}. At the 
network layer, previous work shows that LEO routing is
predictable and the ISL (inter-satellite links) can get congested. ICARUS observes that the constellation
topology is public and deterministic. In this case, a distributed botnet can be used to flood the inter-satellite links and interrupt the communication between two separate regions \cite{giuliari2021icarus}. SKYFALL
demonstrates that time-varying bottleneck links can be identified by an adversary and flooded, causing throughput degradation for legitimate users
\cite{deng2025bottleneck}, and StarMaze constructs periodic, ring-shaped
detours that cause disruption \cite{wang2024starmaze}. There is also existing work for the software
security of the satellites themselves \cite{willbold2023space}.



Existing attacks each have significant limitations. Physical layer attacks such as jamming require specialized equipment and can be detected by the operator. Cyberattacks rely on software or protocol weaknesses, so their effectiveness can be reduced through stronger security controls by the system. Similarly, link-flooding attacks target routing bottlenecks and can be mitigated through randomized routing, traffic rerouting, and bottleneck avoidance \cite{zhang2025isllfa,hu2026pathweaver,meng2025ddos,fratty2023random}. These limitations motivate a different attack approach: adversarial terminals behave like ordinary users, requesting legitimate demand, but collectively manipulating the beam allocation process. Because the attack targets resource allocation rather than network routes, routing defenses alone would not be able to address it. Effective mitigation instead requires robust beam allocation and intra-cell sharing algorithms.

Our contributions are as follows:

\begin{itemize}
\item We introduce and formulate a realistic denial-of-service attack on LEO beam allocation. The adversary purchases and activates additional terminals without breaking the protocol. We model budget on both the number of terminals and targeted cells, and consider attackers that know either the aggregate demand of each cell or only its population. \cref{sec:formulation} formalizes the attack model, adversary knowledge,
and quality-of-service metrics, while \cref{sec:method} presents the attack
heuristics.

\item By evaluating the effect of the heuristics using various metrics, we show how the system configuration changes the effectiveness of the attack strategies. Targeting lightly loaded cells is more effective under equal sharing, while targeting heavily loaded cells is more effective under proportional sharing. Max-min scheduling does not eliminate the attack, but generally requires the adversary to target $1.4\times$ to $2\times$ as many cells as under greedy scheduling. We also find that detailed demand information is useful only when it matches the sharing mechanism; otherwise, simpler population information can perform as well or even better (\cref{sec:eval}).

\item We show that both the intra-cell sharing rule and the beam scheduler affect vulnerability, but in different ways. The sharing rule has a strong effect at every tested budget: at a 5\% budget, proportional sharing causes a 30\% to 34\% decrease in users above 0.7 satisfaction, compared to 9\% to 13\% under equal sharing. The scheduler becomes more important at larger budgets as the injected demand pushes the system toward saturation. We further show that there is a trade-off between robustness and performance, meaning the most robust beam allocation algorithm (max-min equal) has the lowest coverage before the attack happens (\cref{sec:eval}).

\end{itemize}

\section{Background}
\label{sec:background}

This section describes the system where the attack model of \cref{sec:formulation} is evaluated. We first define the constellation and traffic models, then we characterize feasible beam allocations, and finally we describe the beam-allocation algorithms. In this section, the terminal population and its demands are treated as inputs to the beam-allocation algorithms; in \cref{sec:formulation}, we describe how  adversaries can manipulate these inputs.

\subsection{Constellation and Traffic Model}
\label{subsec:constellation}

The service region is partitioned into a set of cells $\mathcal C=\{1,\ldots,C\}$ served by satellites $\mathcal S = \{1,\ldots,S\}$.
At each time~$t$, we denote by $\mathcal S_{i,t} \subseteq\mathcal S$ the set of satellites visible from cell $i \in \mathcal C$. Visibility accounts for line of sight and for the minimum elevation required for a satellite to serve a cell.

Each satellite can activate at most $K$ beams simultaneously, and each cell can receive at most $B$ beams. The downlink spectrum is divided into orthogonal channels $\mathcal F = \{1,\ldots,F\}$: a beam occupies one channel, illuminates one cell, and provides a fixed capacity $W$. The same channel may be used by beams serving sufficiently distant cells; we denote by $\mathcal N \subseteq \mathcal C \times \mathcal C$ the pairs of neighboring cells that cannot use beams with the same channel. Since each beam uses one channel, $B\leq F$.

We denote by $\mathcal U_i = \{1, \dots, U_i\}$ the set of terminals located in cell $i \in \mathcal C$ and refer to $U_i$ as the cell population. We denote by $d_{i,j} \in [0, D_{\max}]$ the service demand of the $j$th terminal in cell $i$, where $D_{\max}$ is the maximum service rate.
The aggregate demand of cell $i$ is
\begin{equation}
    D_i=\sum_{j\in\mathcal U_i}d_{i,j} \quad \text{for each $i\in \mathcal C$}
    \label{eq:aggregate-demand}
\end{equation}
and the network-wide demand vector is $\mathbf D=(D_1,\ldots,D_C)$.

The allocator converts the aggregate demand of each cell into a target number of beams. Because every beam provides capacity $W$, the allocation target of cell $i \in \mathcal C$ is
\begin{equation}
    q_i
    =
    \min\left\{
        \left\lceil\frac{D_i}{W}\right\rceil,
        B
    \right\}.
    \label{eq:allocation-target}
\end{equation}
Thus, $q_i$ is the number of beams needed to cover the cell's aggregate demand, subject to the per-cell beam limit $B$.

\subsection{Beam Allocation Constraints}
\label{subsec:allocation-model}

For each cell $i \in \mathcal C$, satellite $s \in \mathcal S$, and channel $f \in \mathcal F$, we define the binary allocation variable
\begin{equation}
    y_{i,s,f}
    =
    \begin{cases}
        1 & \text{if satellite $s$ serves cell $i$ on channel $f$},\\
        0 & \text{otherwise}.
    \end{cases}
    \label{eq:allocation-variable}
\end{equation}
%
%
The number of beams allocated to cell $i$ is
\begin{equation}
    b_i
    =
    \sum_{s\in\mathcal S}
    \sum_{f\in\mathcal F}
    y_{i,s,f}
    \label{eq:beam-count}
\end{equation}
and the resulting capacity of the cell is $b_iW$. We collect the cell-level allocations in the vector $\mathbf b=(b_1,\ldots,b_C)$.

A feasible beam allocation satisfies the following constraints:
\begin{align}
    &\sum_{i\in\mathcal C}\sum_{f \in \mathcal F}y_{i,s,f}
      \leq K
      &&\forall s\in\mathcal S
      \label{eq:satellite-budget}\\
    &\sum_{s\in\mathcal S}y_{i,s,f}
      \leq 1
      &&\forall i\in\mathcal C,\ \forall f \in \mathcal F
      \label{eq:cell-channel}\\
    &y_{i,s,f}+y_{j,s,f}
      \leq 1
      &&\forall s,\ \forall f,\ \forall (i,j) \in\mathcal N
      \label{eq:adjacent-reuse}\\
    &y_{i,s,f}=0
      &&\forall i,\ \forall f,\ \forall s\notin \mathcal S_{i,t}
      \label{eq:visibility}\\
    &b_i\leq q_i
      &&\forall i\in \mathcal C
      \label{eq:demand-target}
\end{align}

\cref{eq:satellite-budget} enforces the per-satellite beam budget $K$.
\cref{eq:cell-channel} enforces that each cell $i$ can receive a channel $f$ only from one satellite $s$, i.e., that beams assigned to a cell use distinct channels.
\cref{eq:adjacent-reuse} enforces that a satellite must use different channels for neighboring cells.
\cref{eq:visibility} ensures that only visible satellites can serve a cell.
Finally, \cref{eq:demand-target} ensures that the number of assigned beams $b_i$ is lower or equal to $q_i$, the minimum between the allowed number of beams per cell ($B$) and the minimum number of beams necessary to satisfy its demand ($\lceil\frac{D_i}{W}\rceil$).

\subsection{Beam Allocation Algorithms}
\label{subsec:allocation-algorithms}

Given the cell service demands $D_i$, the sets $\mathcal S_{i,t}$ of satellites visible from each cell $i$ at a given time $t$, the sets of neighboring cells $\mathcal N$ and of available frequencies $\mathcal F$, a beam allocation algorithm allocates beams $(i,s,f)$ that satisfy \cref{eq:satellite-budget,eq:cell-channel,eq:adjacent-reuse,eq:visibility,eq:demand-target}.

We consider two beam allocation algorithms \cite{bhosale2025leo}, \textsc{Greedy} and \textsc{Max-Min Fair}.
Both algorithms allocate beams while iterating through cells in order of service demands $D_i$ (starting with higher demands).
For each cell, to assign the next beam among those that satisfy \cref{eq:satellite-budget,eq:cell-channel,eq:adjacent-reuse,eq:visibility,eq:demand-target}, they select one from the least loaded satellite (i.e., the one with the fewest beams assigned so far).
The \textsc{Greedy} algorithm assigns beams in one pass, allocating as many as possible to the current cell (up to its target $q_i$). It therefore gives priority to cells with high aggregate demand.
The \textsc{Max-Min Fair} algorithm allocates beams in multiple passes: during each pass, each cell receives at most one additional beam.
Additional passes are performed until all cells reach their targets $q_i$, or no further feasible beam assignments exist.
This approach prevents a cell from receiving an additional beam before the other cells have been considered, resulting in a fairer distribution of the available beams.

\section{Problem Formulation}
\label{sec:formulation}

In this section, we present the attack strategies used by the adversary to influence beam allocations in a way that reduces service for legitimate users.
We assume that the attacker cannot alter the satellite constellation or the beam allocation algorithm directly, but it can vary the demand of each cell by purchasing additional user subscriptions located in it.
This type of attack can cause degradation of service in two ways: it can change the beam allocation across cells (by altering the service demands $D_i$ and the order in which cells are considered by the algorithms); and it can reduce the share of capacity received by legitimate users inside a cell.
We study how the attacker's knowledge and placement of malicious users affect the service received by legitimate users.

\subsection{Effects of Injection Attacks}
\label{subsec:attack-model}

We denote by $0 \leq b_i(\mathbf a) \leq B$, the number of beams allocated to each cell $i\in \mathcal C$ after an attack $\mathbf a = (a_1, \dots, a_C)$ placing $a_i$~malicious users in cell $i\in \mathcal C$.
To obtain the allocations~$b_i(\mathbf a)$, we assume that each adversarial user injected into the system requests the maximum allowed service demand $D_{\max}$, so that the total demand of cell $i\in \mathcal C$ after the attack becomes $D_i(\mathbf a)=D_i + a_i D_{\max}$ and its beam target becomes $q_i(\mathbf a)=\min\{\lceil\frac{D_i(\mathbf a)}{W}\rceil,B\}$.

By modifying the service demands $D_i(\mathbf a)$, injection of malicious users can change the allocation order of the algorithms, and thus the beams assigned to each cell. Even if injecting additional malicious users does not change the allocation order of a cell, it can still reduce the share of capacity received by users in that cell.
We consider two policies to split capacity among users of a cell: \textsc{Equal} among users or \textsc{Proportional} to their service demands.
Specifically, if we indicate by $s_{i,j} \in [0,d_{i,j}]$ the service demand satisfied for each cell~$i\in \mathcal C$ and user~$j \in \mathcal U_i$,
\begin{align}
    s_{i,j}(\mathbf a)
    =
    \begin{cases}
    \displaystyle \min\left(\frac{b_i(\mathbf a)W}{|\mathcal U_i| + a_i},d_{i,j}\right) & \text{for \textsc{Equal}},\\[5mm]
    \displaystyle \min\left(\frac{b_i(\mathbf a)Wd_{i,j}}{D_i(\mathbf a)},d_{i,j}\right) & \text{for \textsc{Proportional}}.
    \end{cases}
    \label{eq:satisfaction}
\end{align}
Note that the number of injected terminals $a_i$ affects the total cell demands $D_i(\mathbf a)$ and the share of capacity received by legitimate users $\mathcal U_i$ in both cases (by increasing the term at the denominator).
Under \textsc{Equal} sharing, every terminal receives the same fraction $1/(|\mathcal U_i|+a_i)$, so legitimate users with larger demands may have lower fractions of their demands satisfied.
Under \textsc{Proportional} sharing, every terminal receives a share proportional to its demand; consequently, terminals whose service demand is not fully satisfied have the same fraction $b_i(\mathbf a)W/D_i(\mathbf a)$ of satisfied service.

\subsection{Attacker Knowledge}
\label{subsec:attacker-knowledge}

We consider two kinds of attackers:

\paragraph{White-Box Attacker}
The white-box attacker knows the legitimate user demands $d_{i,j}$, the visible satellites $\mathcal S_{i,t}$ from each cell, the beam allocation  algorithm and its constraints, and the capacity sharing policy used in the cells. It can therefore evaluate how a candidate placement of adversarial users will change the demands $D_i(\mathbf a)$, the cell targets $q_i(\mathbf a)$, and the resulting beam allocations $b_i(\mathbf a)$. This situation represents a strong attacker with complete system knowledge, who still cannot modify the allocator or violate the allocation constraints.

\paragraph{Black-Box Attacker}
The black-box attacker has no knowledge of legitimate user cell demands, visible satellites, or details of the allocation algorithm. It can only estimate the number of legitimate users in a cell from public information on its population; in addition, for its injected terminals, it knows their locations, nominal demands $d_{i,j}$ and satisfied demands $s_{i,j}$.
The attacker may combine observations from the terminals it controls across cells and snapshots, but it cannot observe the individual demands or service rates of other users. It may therefore only rank candidate cells using the number of terminals instead of their actual service demands.

\smallskip
Both attackers have the same budgets $A_{\max}$ for the number of injected terminals and $T_{\max}$ for the total number of attacked cells, i.e.,
$
\lVert\mathbf a\rVert_1 \leq A_{\max}
\text{ and }
\lVert\mathbf a\rVert_0 \leq T_{\max}
$.
The knowledge model changes how the placement is
chosen, but not its feasible set or its effect on demand. A small $T_{\max}$
forces the attack to concentrate terminals in a few cells, whereas a larger
$T_{\max}$ permits the same terminal budget to be distributed more widely.

\subsection{Quality of Service Metrics}
\label{subsec:service-metrics}

Let $\mathcal U = \cup_{i\in \mathcal C} \,\mathcal U_i$ denote the set of all legitimate users. We evaluate three metrics for the quality of service:
\begin{align}
    \overline S(\mathbf a)
    &=
    \frac{1}{|\mathcal U|}
    \sum_{i \in \mathcal C}
    \sum_{j\in\mathcal U_i}
    \frac{s_{i,j}(\mathbf a)}{d_{i,j}}\,,
    \label{eq:average-satisfaction}\\
    S_\theta(\mathbf a)
    &=
    \frac{1}{|\mathcal U|}
    \sum_{i \in \mathcal C}
    \sum_{j\in\mathcal U_i}
    \mathbf 1
    \left[\frac{s_{i,j}(\mathbf a)}{d_{i,j}}\geq\theta\right]\,,
    \label{eq:below-threshold}\\
    \mathrm{Cov}(\mathbf a)
    &=
    \frac{
      \sum_{i \in\mathcal C}
      \sum_{j\in\mathcal U_i}
        s_{i,j}(\mathbf a)
    }{
      \sum_{i \in\mathcal C}
      \sum_{j\in\mathcal U_i}
        d_{i,j}
    }.
    \label{eq:coverage}
\end{align}
The metric $\overline S$ represents the average fraction of satisfied demand among legitimate users; $S_\theta$ is the fraction of legitimate users with at least $\theta$\% of their demands satisfied; and $\mathrm{Cov}$ is the fraction of total demand of legitimate users that is satisfied.
These metrics are complementary: $\overline S$ treats all legitimate users equally, $S_\theta$ captures legitimate users with a minimum quality of service, and $\mathrm{Cov}$ measures the total fraction of legitimate demand served.
Average satisfaction summarizes overall degradation but may hide a small group of severely affected users. The threshold metric of $S_\theta$ exposes this lower tail, while $\mathrm{Cov}$ measures how much requested legitimate traffic the system actually serves.

\subsection{Adversarial Objectives}
\label{subsec:objectives}

We consider attacks over multiple time points $\mathcal T$ and denote at time point $t\in \mathcal T$ as an additional argument, i.e.,  for each metric $M\in\{\overline S,S_\theta,\mathrm{Cov}\}$, $M(\mathbf a, t)$ is the metric a time $t$ after attack $\mathbf a$.
Quality of service degradation $$\delta M(\mathbf a;t) = M(\mathbf 0;t) - M(\mathbf a;t)$$ is the difference of a quality of service metric before and after the attack.
Note that the time point $t\in \mathcal T$ determines the satellites $\mathcal S_{i,t}$ visible from each cell $i\in C$, but we assume that the attacker cannot move injected terminals over time.

For the attacker, the objective is the average service degradation over time. Specifically, for each metric $M\in\{\overline S,S_\theta,\mathrm{Cov}\}$ the objective is
\begin{align}
\delta M(\mathbf a)
=
\frac{1}{|\mathcal T|}
\sum_{t\in\mathcal T} \delta M(\mathbf a; t)
\end{align}
and an optimal attack strategy  is
\begin{equation}
    \mathbf a_M^\star
    \in
    \operatorname*{arg\,max}\limits_{\substack{\lVert\mathbf a\rVert_1 \leq A_{\max}\\\lVert\mathbf a\rVert_0 \leq T_{\max}}}
    \delta M(\mathbf a)\,.
    \label{eq:attack-objective}
\end{equation}
White-box and black-box attacks use the same objectives and budgets; they differ only in the information available when selecting $\mathbf a$. In the evaluation, we compare placement heuristics appropriate to each knowledge model. White-box heuristics may use internal demand and allocation information, whereas black-box heuristics may use only terminal-side measurements. Because the allocation map is discrete and coupled through the beam constraints, we compare the degradation achieved by these heuristics instead of assuming that a closed-form optimum is available.

\begin{table*}[t]
\centering
\caption{Most damaging heuristic per budget, mean demand 30~Mbps ($\approx80\%$ utilization). Value $\pm$ 95\% CI; superscript is the winning heuristic (b/t/r = bottom/top/random). \dashuline{Dashed}: best across all algorithms. Lower is worse for users.}
\label{tab:summary-mean30}
\scriptsize
\setlength{\tabcolsep}{3pt}
\begin{tabular}{|l|l|c|ccccc|}
\hline
\multirow{2}{*}{Scheduler / capacity sharing} & \multirow{2}{*}{Metric} & \multirow{2}{*}{0\%} & \multicolumn{5}{c|}{Adversary budget} \\
\cline{4-8}
 & & & 1\% & 2\% & 3\% & 4\% & 5\% \\
\hline
\multirow{3}{*}{Greedy, equal} & Coverage & $61.5$ & $61.1\,{\scriptstyle\pm0.1}^{\,b}$ & $59.4\,{\scriptstyle\pm0.3}^{\,b}$ & $56.8\,{\scriptstyle\pm0.4}^{\,b}$ & $53.7\,{\scriptstyle\pm0.5}^{\,b}$ & $49.6\,{\scriptstyle\pm0.6}^{\,b}$ \\
\cline{2-8}
 & Mean satisfaction & $73.9$ & $73.5\,{\scriptstyle\pm0.1}^{\,b}$ & $71.7\,{\scriptstyle\pm0.3}^{\,b}$ & $69.1\,{\scriptstyle\pm0.4}^{\,b}$ & $65.7\,{\scriptstyle\pm0.5}^{\,b}$ & $61.6\,{\scriptstyle\pm0.7}^{\,b}$ \\
\cline{2-8}
 & Over 0.7 & $62.8$ & $62.3\,{\scriptstyle\pm0.1}^{\,t}$ & $60.6\,{\scriptstyle\pm0.3}^{\,b}$ & $57.9\,{\scriptstyle\pm0.4}^{\,b}$ & $54.3\,{\scriptstyle\pm0.5}^{\,b}$ & $49.7\,{\scriptstyle\pm0.7}^{\,b}$ \\
\hline
\multirow{3}{*}{Greedy, proportional} & Coverage & $73.6$ & $69.7\,{\scriptstyle\pm0.1}^{\,t}$ & $67.1\,{\scriptstyle\pm0.1}^{\,t}$ & $64.9\,{\scriptstyle\pm0.1}^{\,t}$ & $62.9\,{\scriptstyle\pm0.1}^{\,t}$ & $59.8\,{\scriptstyle\pm0.8}^{\,b}$ \\
\cline{2-8}
 & Mean satisfaction & $73.6$ & $69.7\,{\scriptstyle\pm0.0}^{\,t}$ & $67.1\,{\scriptstyle\pm0.0}^{\,t}$ & $64.9\,{\scriptstyle\pm0.0}^{\,t}$ & $62.9\,{\scriptstyle\pm0.0}^{\,t}$ & $59.7\,{\scriptstyle\pm0.8}^{\,b}$ \\
\cline{2-8}
 & Over 0.7 & $62.5$ & $46.8\,{\scriptstyle\pm0.2}^{\,t}$ & $40.5\,{\scriptstyle\pm0.2}^{\,t}$ & $35.7\,{\scriptstyle\pm0.2}^{\,t}$ & $31.4\,{\scriptstyle\pm0.2}^{\,t}$ & $28.1\,{\scriptstyle\pm0.2}^{\,t}$ \\
\hline
\multirow{3}{*}{Max-min, equal} & Coverage & $56.8$ & \dashuline{$55.6\,{\scriptstyle\pm0.4}^{\,b}$} & \dashuline{$53.6\,{\scriptstyle\pm0.1}^{\,b}$} & \dashuline{$52.1\,{\scriptstyle\pm0.6}^{\,b}$} & \dashuline{$50.2\,{\scriptstyle\pm0.6}^{\,b}$} & \dashuline{$49.2\,{\scriptstyle\pm0.2}^{\,b}$} \\
\cline{2-8}
 & Mean satisfaction & $69.0$ & $68.0\,{\scriptstyle\pm0.4}^{\,b}$ & $65.7\,{\scriptstyle\pm0.1}^{\,b}$ & $64.4\,{\scriptstyle\pm0.6}^{\,b}$ & $62.3\,{\scriptstyle\pm0.7}^{\,b}$ & $61.2\,{\scriptstyle\pm0.2}^{\,b}$ \\
\cline{2-8}
 & Over 0.7 & $56.8$ & $55.5\,{\scriptstyle\pm0.5}^{\,b}$ & $52.8\,{\scriptstyle\pm0.2}^{\,b}$ & $51.0\,{\scriptstyle\pm0.7}^{\,b}$ & $48.6\,{\scriptstyle\pm0.7}^{\,b}$ & $47.4\,{\scriptstyle\pm0.3}^{\,b}$ \\
\hline
\multirow{3}{*}{Max-min, proportional} & Coverage & $67.2$ & $63.3\,{\scriptstyle\pm0.1}^{\,t}$ & $60.6\,{\scriptstyle\pm0.1}^{\,t}$ & $58.2\,{\scriptstyle\pm0.1}^{\,t}$ & $56.2\,{\scriptstyle\pm0.1}^{\,t}$ & $54.3\,{\scriptstyle\pm0.1}^{\,t}$ \\
\cline{2-8}
 & Mean satisfaction & $67.2$ & \dashuline{$63.3\,{\scriptstyle\pm0.1}^{\,t}$} & \dashuline{$60.6\,{\scriptstyle\pm0.1}^{\,t}$} & \dashuline{$58.3\,{\scriptstyle\pm0.1}^{\,t}$} & \dashuline{$56.2\,{\scriptstyle\pm0.1}^{\,t}$} & \dashuline{$54.3\,{\scriptstyle\pm0.1}^{\,t}$} \\
\cline{2-8}
 & Over 0.7 & $53.1$ & \dashuline{$39.4\,{\scriptstyle\pm0.4}^{\,t}$} & \dashuline{$32.4\,{\scriptstyle\pm0.3}^{\,t}$} & \dashuline{$27.9\,{\scriptstyle\pm0.4}^{\,t}$} & \dashuline{$24.6\,{\scriptstyle\pm0.5}^{\,t}$} & \dashuline{$22.0\,{\scriptstyle\pm0.5}^{\,t}$} \\
\hline
\end{tabular}
\end{table*}

\section{Attack Heuristics}
\label{sec:method}

In real-world network deployments, solving the attack-placement problem exactly can require prohibitive computation time because the allocation variables are discrete and the beam constraints couple decisions across cells. We therefore use computationally efficient heuristics derived from a single system snapshot. Each heuristic has two components: a \emph{target selection}, which selects a set $\mathcal V\subseteq\mathcal C$ of cells with $|\mathcal V|\leq T_{\max}$; and a \emph{spreading rule}, which distributes the $A_{\max}$ injected terminals among the selected cells. Once computed, the placement $\mathbf a$ remains fixed throughout $\mathcal T$.

\subsection{Target Selection}
\label{subsec:target-selection}

Each heuristic assigns a score $\kappa_i$ to each cell $i\in \mathcal C$.
The score depends on the attacker's knowledge: a white-box attacker uses the aggregate demand $\kappa_i=D_i$, because demand determines both the allocation order and the beam targets $q_i$; a black-box attacker observes only the cell population and therefore uses $\kappa_i=|\mathcal U_i|$ as a proxy for aggregate cell demand.

Let $\pi$ be a permutation of the cells $\mathcal C = \{1,\dots, C\}$ such that $\kappa_{\pi(1)}\leq\cdots\leq\kappa_{\pi(C)}$, and define $1\le k\le\min\{T_{\max},C\}$. We consider three target selection rules:
\begin{equation}
    \mathcal V =
    \begin{cases}
        \{\pi(1),\ldots,\pi(k)\}
            & \text{for \textsc{Bottom},}\\
        \{\pi(C),\ldots,\pi(C-k+1)\}
            & \text{for \textsc{Top},}\\
        \mathrm{Rand}_k(\mathcal C)
            & \text{for \textsc{Random},}
    \end{cases}
    \label{eq:target-set}
\end{equation}
where $\mathrm{Rand}_k(\mathcal C)$ denotes a $k$-element subset of $\mathcal C$ selected uniformly at random.

The heuristic \textsc{Top} selects cells with the largest scores $\kappa_i$. These cells are typically assigned beams before other cells, and diverting  resources from them can affect many legitimate users. Their beam targets $q_i$ are often close to the per-cell maximum~$B$: additional demand may  leave target $q_i$ unchanged, while reducing the capacity share of legitimate users.

The heuristic \textsc{Bottom} selects cells with the smallest scores, which are typically assigned beams after other cells. In these cells, even injecting a small number of adversarial users can increase  $D_i$ and move the cell earlier in the allocation order.

Finally, the heuristic \textsc{Random} provides a baseline for measuring the benefit of ranking information.

\subsection{Spreading Rules}
\label{subsec:spreading}

Let $(v_1,\ldots,v_k)$ denote the ordering of the selected cells $\mathcal V$ according to \cref{eq:target-set} (i.e., by increasing or decreasing score $\kappa_i$ for \textsc{Bottom} or \textsc{Top}, and a random order for \textsc{Random}).
The attacks assign $a_i=0$ for every cell $i\notin\mathcal V$; for selected cells $\mathcal V$, two strategies are considered to assign injected terminals:

\paragraph{Uniform Spreading}
The terminal budget is divided as evenly as possible across the selected cells $\mathcal V$, irrespective of their scores $\kappa_i$.
Specifically, for $i=1,\dots,k$,
\begin{equation}
    a_{v_i}
    =
    \begin{cases}
    \left\lceil\frac{A_{\max}}{k}\right\rceil
     & \text{if $i\leq A_{\max}\bmod k$,}\\
    \left\lfloor\frac{A_{\max}}{k}\right\rfloor & \text{otherwise.}
    \end{cases}
    \label{eq:spread-uniform}
\end{equation}
Thus, the first $A_{\max}\bmod k$ cells receive one additional terminal and $\sum_{i=1}^k a_{v_i}=A_{\max}$.

\paragraph{Water-Filling}
Water-filling distributes the injected terminals among selected cells $\mathcal V$ according to the scores $\kappa_i$ (i.e., the aggregate cell demand for a white-box attacker, or the cell population for a black-box attacker).
Each cell $i \in \mathcal V$ starts with $a_i = 0$.
At each step, an additional injected terminal is assigned to the cell $i \in \mathcal V$ with the smallest $\kappa_i+\delta a_i$ (with ties broken arbitrarily), where $\delta=D_{\max}$ for a white-box attacker and  $\delta=1$ for a black-box attacker.
This process is repeated for $A_{\max}$ steps, to assign all terminals.

\smallskip
Thus, while uniform spreading attempts to balance the number of injected terminals, water-filling attempts to balance the scores $\kappa_i+\delta a_i$.
For a white-box attacker ranking cells based on their aggregate demands, water-filling produces similar aggregate demands $D_i(\mathbf a)$ and beam targets $q_i(\mathbf a)$.
For a black-box attacker ranking cells based on their populations, water-filling produces similar cell populations $|\mathcal U_i|+a_i$ after the attack, which is the sharing denominator in \cref{eq:satisfaction} for \textsc{Equal} sharing.

Cells that already have high $\kappa_i$ may receive no injected terminals with water-filling: the number $k$ of selected cells is only an upper bound on the number of cells that actually receive injected terminals. Under uniform spreading, if $A_{\max}<k$, then $k-A_{\max}$ selected cells receive no terminals. Note that both spreading rules assign the entire $A_{\max}$ budget.

\section{Experimental Evaluation}
\label{sec:eval}

\subsection{Setup}
\paragraph{Simulation Platform and Constellation}
We evaluate our attacks on CosmoSim \cite{bhosale2025leo}, a discrete LEO
satellite network simulator that models beam allocation and spectrum reuse
over a time-varying constellation topology. The constellation consists of
$S=6,364$ satellites, each with $K=32$ beams and $F=8$ channels. Thus,
each channel can be reused by at most four beams per satellite. The orbital
periods range from $95.24$ to $96.06$ minutes with a mean of approximately
$95.6$ minutes.

The topology is sampled at $|\mathcal T|=95$ snapshots spaced $60$ seconds
apart, spanning $94$ minutes. This period is nearly one complete orbit of
the constellation, which is sufficient to capture the time-varying visibility
of satellites from each cell.

\paragraph{Service Area and Demand Model}
The service area is Great Britain, discretized into $943$ H3 hexagonal cells.
We deploy $U=100,000$ legitimate users, distributed across cells proportionally to their populations.

Each satellite provides a Ku-band downlink of $W=1,280$ Mbps per beam, with
eight channels per cell and a frequency reuse factor of four, giving an
aggregate downlink capacity of $KW=40,960$ Mbps per satellite. At the
reference epoch (snapshot $t=0$), $|S_{0}|=92$ satellites are simultaneously visible
from the service area, providing a total downlink capacity of approximately
$C_T=KW|S_{0}|=3.77$ Tbps. We assume that all terminals are capable of receiving the
maximum downlink rate $D_{\max}=200$ Mbps. Our experiments only focus on user
downlink capacity.

Each legitimate user $j$ in cell $i$ has a service demand $d_{i,j}$ drawn
from a truncated log-normal \cite{alasmar2021lognormal} distribution
with mean $\mu$ and a coefficient of variation of $0.67$, truncated to the
interval $[0,D_{\max}]$. We define the utilization factor $\rho$ as the
ratio of the total demand to the total capacity $C_T=KW|S_{0}|$ of the visible satellites:

\begin{equation}
    \rho = \frac{\sum_{i\in\mathcal C}\sum_{j\in\mathcal U_i} d_{i,j}}{C_T}.
    \label{eq:utilization}
\end{equation}

The mean $\mu$ is set to $\rho C_T/U$, where $U$ is the total number of legitimate users.
At the default utilization factor $\rho=0.8$, the total demand is approximately
$3$ Tbps, which yields an average demand of $\mu=30$ Mbps per user over a
heavy-tailed distribution. We also evaluate the system under a lower utilization
factor $\rho=0.6$ and a higher utilization factor $\rho=1.0$, which correspond to
average demands of $\mu=22.5$ Mbps and $\mu=37.5$ Mbps, respectively.

\paragraph{Resource Allocation Policies}
As mentioned in \cref{subsec:allocation-algorithms} and \cref{subsec:attack-model},
we evaluate two beam allocation algorithms, \textsc{Greedy} and \textsc{Max-Min Fair},
and two intra-cell capacity sharing policies, \textsc{Equal} and \textsc{Proportional}.
The combination of these policies results in four resource allocation policies:
\textsc{Greedy-Equal}, \textsc{Greedy-Proportional}, \textsc{Max-Min Fair-Equal},
and \textsc{Max-Min Fair-Proportional}.

\paragraph{Threat Model and Attack Strategies}
We assume two types of attackers, as specified in \cref{subsec:attacker-knowledge}:
a white-box attacker with complete knowledge of the system and a black-box attacker
with limited knowledge. Both attackers have the same budgets for the number of injected
terminals $A_{\max}$ and the number of attacked cells $T_{\max}$. Adversarial terminals
are assumed to request the maximum service demand $D_{\max}$, which is $200$ Mbps in our experiments.

In our experiments, we set $T_{\max}=300$ and vary $A_{\max}$ from $1,000$ to $5,000$,
which corresponds to $1\%$ to $5\%$ of the total legitimate user population.
However, the number of targeted cells $k$ is not constant and varies up to $T_{\max}$, and is
optimized by the attacker to maximize the impact of the attack.

We evaluate three attack heuristics, as described in \cref{sec:method}:
\textsc{Bottom}, \textsc{Top}, and \textsc{Random}, combined with two
spreading rules, \textsc{Uniform} and \textsc{Water-Filling}, resulting
in six attack strategies.

\paragraph{Metrics}
We report three complementary metrics as stated in \cref{subsec:service-metrics}:
average satisfaction $\overline S$, the fraction of users with at least $70\%$
of their demand satisfied $S_{0.7}$, and coverage $\mathrm{Cov}$, all computed
over legitimate terminals only, with adversary terminals excluded so that the
attacker's own inflated demand cannot flatter the results.

\paragraph{Experimental Design and Statistical Treatment}
Every deterministic configuration is repeated across 20 independent random
demand realizations (drawn with different random seeds), while the other parameters
are held fixed. Reported values are means over these realizations, and error bars
are $95\%$ confidence intervals computed as $\pm 1.96 \times \text{SEM}$. For randomly selected configurations, the same 20 demand
realizations are used, and the 10 cell-selection samples within each
realization are averaged before aggregation. This ensures that the
reported confidence intervals reflect variability across demand
realizations, consistent with the other heuristics, rather than
variability due to random cell-selection draws.

In the persistent experiment, the attack is planned once, against the constellation
at the reference epoch, and replayed unchanged at every subsequent snapshot.
Each snapshot is repeated across twenty independent demand realizations, and the same
confidence intervals as above are computed over the twenty realizations, with the
twenty demand realizations remaining fixed across snapshots so that the only variation
is the constellation's visibility. The utilization factor $\rho$ is left to
drift as the visible constellation changes size: subscribers do not revise their
demand because a satellite moved, and absorbing capacity that comes and goes
is what a persistent attack must survive.

Where we report the effect of a single parameter, configurations are matched on all
remaining parameters and compared pairwise, rather than by differencing marginal means,
so that an imbalance elsewhere in the design cannot be mistaken for the effect
under study. Where we report worst-case vulnerability, we take the maximum over attack
strategies rather than the mean, since an adversary chooses their strategy and would
not adopt an average one.

\begin{figure}[t]
\centering
\begin{subfigure}[t]{0.48\columnwidth}
\centering
\includegraphics[width=\linewidth]{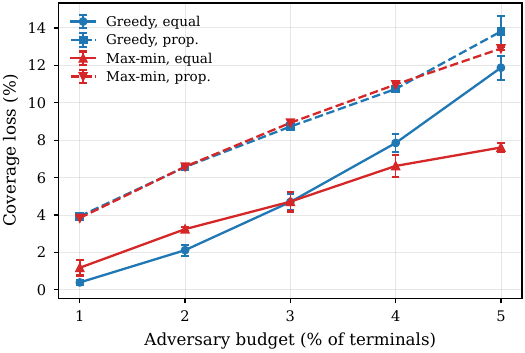}
\caption{Coverage loss.}
\label{fig:coverage-summary}
\end{subfigure}\hfill
\begin{subfigure}[t]{0.48\columnwidth}
\centering
\includegraphics[width=\linewidth]{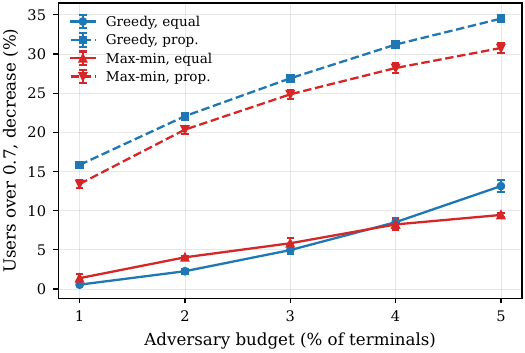}
\caption{Decrease in users at/above 0.7 satisfaction.}
\label{fig:over70}
\end{subfigure}
\caption{Worst-case user harm versus adversary budget at mean demand 30~Mbps. Each point selects the most damaging tested attack for that scheduler/sharing configuration; error bars show 95\% confidence intervals. Proportional sharing (dashed) causes similar coverage and mean-satisfaction losses under the two schedulers but a much larger decrease in the share of users at/above 0.7 than equal sharing (solid).}
\label{fig:summary-mean30}
\end{figure}

\subsection{Main Results}
\label{sec:results}

The results reveal three main patterns. First, the intra-cell sharing rule has a particularly strong effect on vulnerability: proportional sharing generally permits substantially greater damage than equal sharing, especially to the fraction of users getting at least 0.7 demand satisfaction. The scheduler nonetheless remains important, and its influence becomes more pronounced at larger adversary budgets, as the additional adversarial demand increases system load. Second, robustness comes at a cost. Max-min/equal experiences the smallest attack-induced degradation, but it also has the lowest coverage before any attack; its smaller loss therefore reflects, in part, a weaker starting point rather than uniformly better service. Third, the attacker's most effective targeting strategy depends on the resource allocation policies: targeting the least loaded cells is generally most damaging under equal sharing, whereas targeting the most-loaded cells is most damaging under proportional sharing. Table~\ref{tab:summary-mean30} and Fig.~\ref{fig:summary-mean30} summarize these patterns across every scheduler sharing configuration and all three user-level metrics, reporting the maximum damage achieved by the tested heuristics at each adversary budget. Furthermore, we will discuss the effect of two important choices in the heuristics, which are the spread rules (water-fill or uniform) and the number of cells to attack.

\subsubsection{Effects of resource allocation policies}
Both the intra-cell sharing rule and the beam scheduler shape the impact of an attack, although the sharing rule produces the larger separation overall. This difference is most pronounced for the share of users getting at least 0.7 satisfaction. At a 5\% adversary budget, this share decreases by 34 percentage points under greedy/proportional and 30 points under max-min/proportional, compared with 13 and 9 points under the corresponding equal-sharing configurations (Table~\ref{tab:summary-mean30}; Fig.~\ref{fig:over70}). Thus, proportional sharing substantially amplifies the attacker's leverage under either scheduler. The scheduler nevertheless affects how the system responds as the adversary budget increases the offered load. At the 5\% budget, the loss in users getting at least 0.7 satisfaction is 3.7 points greater under greedy than under max-min with equal sharing, and 3.3 points greater with proportional sharing; the same high-budget effect appears in coverage.
\begin{figure}[t]
\centering
\begin{subfigure}[t]{0.48\columnwidth}
\centering
\includegraphics[width=\linewidth]{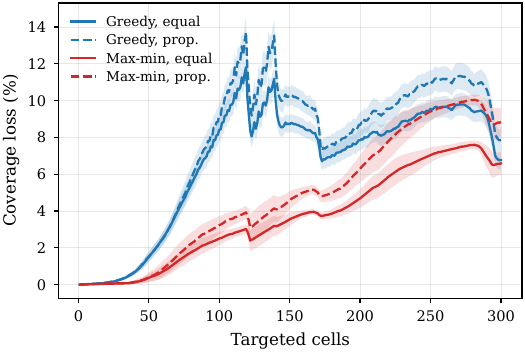}
\caption{Bottom heuristic.}
\label{fig:cells-coverage-bottom}
\end{subfigure}\hfill
\begin{subfigure}[t]{0.48\columnwidth}
\centering
\includegraphics[width=\linewidth]{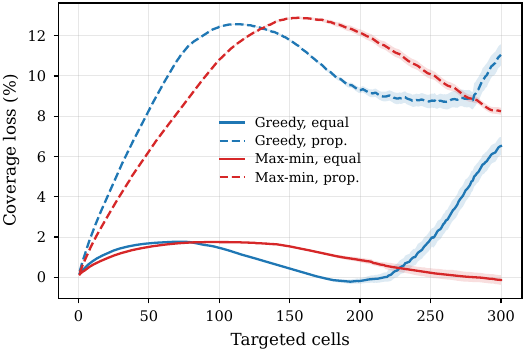}
\caption{Top heuristic.}
\label{fig:cells-coverage-top}
\end{subfigure}
\caption{Effect of the number of demand-ranked target cells on coverage loss, at a representative 5\% adversary budget. Lines show the mean and shaded regions show uncertainty. Both the maximizing width and the shape of the response depend on the scheduler, sharing rule, and targeting heuristic.}
\label{fig:cells-coverage}
\end{figure}

\begin{figure*}[t]
\centering
\begin{subfigure}[t]{0.24\textwidth}
\centering
\includegraphics[width=\linewidth]{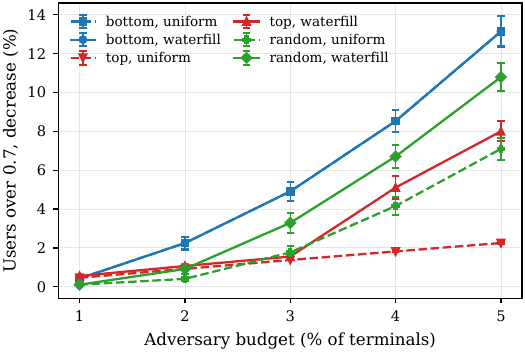}
\caption{Greedy scheduler, equal sharing.}
\label{fig:over70-greedy-equal}
\end{subfigure}\hfill
\begin{subfigure}[t]{0.24\textwidth}
\centering
\includegraphics[width=\linewidth]{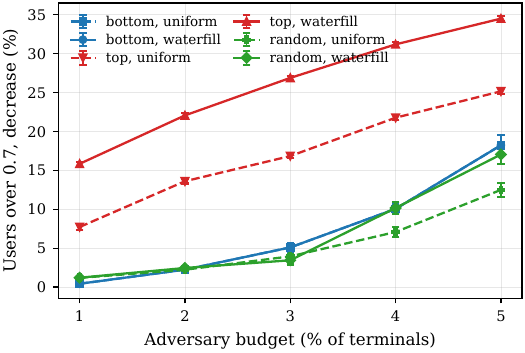}
\caption{Greedy scheduler, proportional sharing.}
\label{fig:waterfill}
\end{subfigure}\hfill
\begin{subfigure}[t]{0.24\textwidth}
\centering
\includegraphics[width=\linewidth]{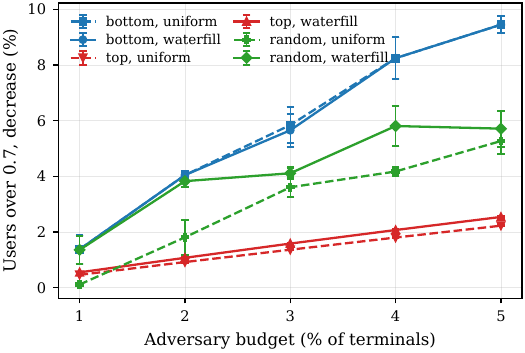}
\caption{Max-min scheduler, equal sharing.}
\label{fig:over70-maxmin-equal}
\end{subfigure}\hfill
\begin{subfigure}[t]{0.24\textwidth}
\centering
\includegraphics[width=\linewidth]{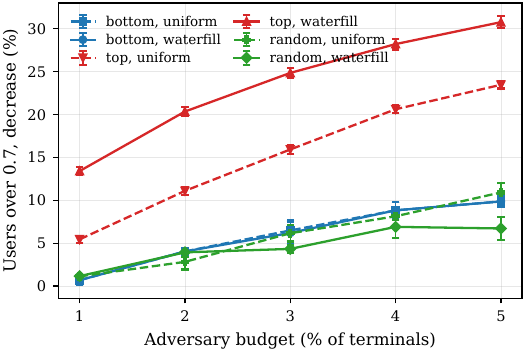}
\caption{Max-min scheduler, proportional sharing.}
\label{fig:over70-maxmin-prop}
\end{subfigure}
\caption{Decrease in the share of users at/above 0.7 satisfaction by adversary budget for every heuristic and spread rule, with demand-ranked cells. For top targeting under proportional sharing, waterfill (solid red) opens a growing lead over uniform spreading (dashed red); bottom and random targeting are much less sensitive to the spread rule.}
\label{fig:over70-heuristics}
\end{figure*}

\subsubsection{Robustness and efficiency trade-off}
With no attack, max-min/equal has the lowest baseline coverage of the four configurations (56.8\%, against 61.5\% for greedy/equal and 73.6\%/67.2\% for the proportional pair) and also the smallest worst-case coverage drop at 5\% budget (7.6 points, against 11.9 to 13.8 for the others, as we can see in Table~\ref{tab:summary-mean30}). Max-min/equal combines two mechanisms that limit an attacker: beam spread evenly across cells and capacity split evenly within a cell. Both also limit how well it serves demand with no attack present. It is the safe choice only for an operator who already accepts low coverage; a drop measured from a lower starting point is not automatically a better outcome.

\subsubsection{Most effective heuristic}
\begin{figure*}[t]
\centering
\begin{subfigure}[t]{0.24\textwidth}
\centering
\includegraphics[width=\linewidth]{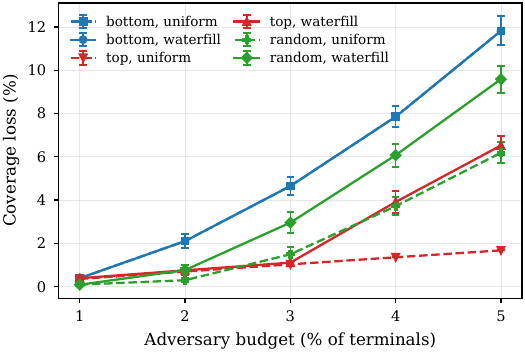}
\caption{Greedy scheduler, equal sharing.}
\label{fig:coverage-greedy-equal}
\end{subfigure}\hfill
\begin{subfigure}[t]{0.24\textwidth}
\centering
\includegraphics[width=\linewidth]{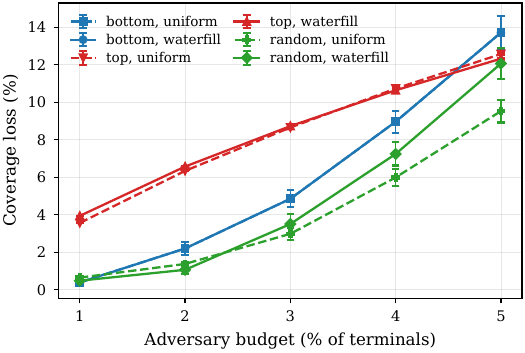}
\caption{Greedy scheduler, proportional sharing.}
\label{fig:coverage-greedy-prop}
\end{subfigure}\hfill
\begin{subfigure}[t]{0.24\textwidth}
\centering
\includegraphics[width=\linewidth]{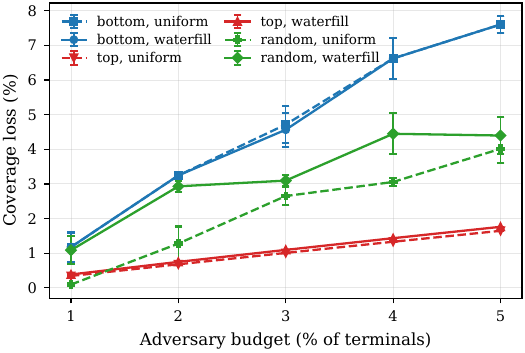}
\caption{Max-min scheduler, equal sharing.}
\label{fig:coverage-maxmin-equal}
\end{subfigure}\hfill
\begin{subfigure}[t]{0.24\textwidth}
\centering
\includegraphics[width=\linewidth]{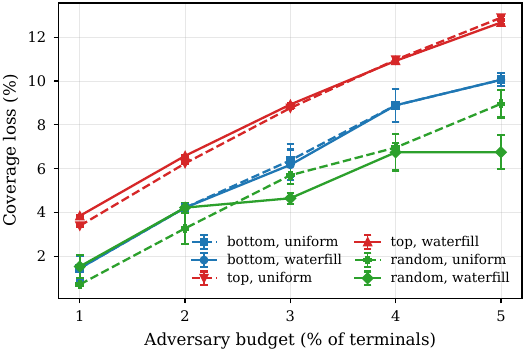}
\caption{Max-min scheduler, proportional sharing.}
\label{fig:coverage-maxmin-prop}
\end{subfigure}
\caption{Coverage loss by adversary budget for every cell-selection heuristic and spread rule, with demand-ranked cells. Bottom targeting is strongest under equal sharing, whereas top targeting becomes competitive under proportional sharing. Solid and dashed versions of the same color compare waterfill and uniform spreading.}
\label{fig:coverage-heuristics}
\end{figure*}

The most damaging tested heuristic is determined primarily by the intra-cell sharing rule. Under equal sharing, targeting the least-loaded cells (bottom) wins in 29 of the 30 scheduler, metric, budget combinations reported in Table~\ref{tab:summary-mean30}. The only exception occurs for greedy/equal on the fraction of users getting at least 0.7 satisfaction at the 1\% budget, which is very close to other heuristics, as we can see in Figure~\ref{fig:over70-greedy-equal}. Under proportional sharing, this pattern reverses: targeting the most-loaded cells (top) wins in 28 of the 30 combinations. Because capacity is divided in proportion to demand, placing high-demand adversarial users in already heavily loaded cells displaces a larger amount of legitimate demand. The exceptions are coverage and mean satisfaction under greedy/proportional at the 5\% budget. At this load, the most heavily loaded cells have already reached their beam caps, limiting the marginal damage caused by additional adversarial demand; targeting less-loaded cells therefore becomes more effective. Figures~\ref{fig:coverage-heuristics} and~\ref{fig:over70-heuristics} show a broadly consistent pattern across the two metrics: bottom targeting is most effective under equal sharing, while top targeting is generally strongest under proportional sharing. The main difference is the magnitude of the effect. Top targeting creates a much larger separation in the fraction of users attaining at least 0.7 satisfaction, whereas bottom targeting can approach or exceed it for coverage at the highest budget. Random targeting consistently produces less damage than the targeted heuristics. Bottom and random targeting also have wider confidence intervals because their selected cells are more sensitive to demand realizations. Small demand changes can reorder the least-loaded cells or change the effectiveness of a random target set, while the heavily loaded cells selected by top targeting remain comparatively stable.

\subsubsection{Number of attacked cells}
The best number of targeted cells depends heavily on the beam allocation scheduler, and more cells must be attacked for max-min. For the share of users above 0.7 at 5\% budget, the best attack uses 246 cells under max-min/proportional against 180 under greedy/proportional (from 126 vs.\ 88 at 1\% budget), roughly $1.4\times$ throughout. Under equal sharing, the gap is wider: 286 cells for max-min against 139 for greedy at 5\% budget, close to $2\times$. Because max-min hands out beams one at a time and keeps allocations level, pressure on one cell does not starve its neighbours the way it does under greedy, so the attacker must widen the attack rather than deepen it. The fixed budget response curves in Figure~\ref{fig:cells-coverage} show why the best width must be optimized separately for each design. Bottom targeting produces several local peaks as additional cells cross allocation thresholds, while top targeting under proportional sharing has a clear scheduler-dependent interior maximum before the attack becomes too diffuse.

\subsubsection{Spread rule effect}
Once the target cells are selected, the adversary can distribute its budget uniformly or use water-filling to balance the post-attack ranking scores. Table~\ref{tab:heuristics-mean30} compares these spread rules at the 5\% budget for both demand and user based rankings. The main finding is that water-filling is not inherently more damaging; its effectiveness depends on whether the quantity it balances matches the source of congestion.

For bottom targeting, uniform spreading and water-filling produce nearly identical results across schedulers, sharing rules, and ranking bases. Because the selected cells are all comparatively lightly loaded, there is little initial imbalance for water-filling to correct. Consequently, the two rules impose similar pressure on the selected cells, with differences below one percentage point.

The spread rule matters most for top targeting under proportional sharing, where its effect reverses with the ranking basis, as we can see in Figure~\ref{fig:over70}. With demand-based ranking, water-filling avoids spending additional budget on cells that have already reached their beam caps and redirects it toward the next-most-loaded cells, where additional demand can still displace legitimate traffic. For example, under greedy/proportional, the decrease in users attaining at least 0.7 satisfaction grows from 25.2\% with uniform spreading to 34.5\% with water-filling; Figure~\ref{fig:waterfill} shows this separation emerging from the lowest budget. With user-based ranking, however, water-filling balances population rather than the aggregate demand that governs proportional sharing. This mismatch shifts budget away from the most constrained cells and reduces damage, from 25.4\% to 16.1\% in the same greedy configuration. Max-min exhibits the same qualitative reversal, and the coverage results follow a similar pattern.

Under equal sharing, the spread rule generally has much less influence because capacity depends on the number of users rather than their individual demands. The main exception is greedy with demand-based top targeting, where water-filling produces a noticeable increase in damage, although the absolute loss remains substantially smaller than under proportional sharing. Overall, water-filling is most effective when it avoids saturated cells while balancing a score that accurately represents the system's true load; when the ranking basis is misaligned with the sharing rule, it can instead weaken the attack. Figures~\ref{fig:coverage-heuristics} and~\ref{fig:over70-heuristics} show that, for random targeting, water-filling generally produces greater damage than uniform spreading, except under max-min/proportional. Water-filling usually improves random targeting by compensating for initial load differences among the randomly selected cells. Under max-min/proportional, however, the scheduler distributes beams more evenly across cells. Uniform spreading therefore preserves a wider attack footprint, whereas water-filling may concentrate the injected terminals in fewer cells. As a result, uniform spreading is more effective in this configuration.

\begin{table}[t]
\centering
\caption{Worst-case attack delta (\%) by utilization; 60\% = 22.5~Mbps mean demand, 80\% = 30~Mbps, 100\% = 37.5~Mbps. Maximized over every attack configuration and budget. Value $\pm$ 95\% CI. \dashuline{Dashed}: lowest absolute value at that utilization. Higher is worse for users.}
\label{tab:utilization}
\scriptsize
\setlength{\tabcolsep}{3pt}
\begin{tabular}{|l|l|ccc|}
\hline
\multirow{2}{*}{Scheduler / capacity sharing} & \multirow{2}{*}{Metric} & \multicolumn{3}{c|}{Utilization} \\
\cline{3-5}
 & & 60\% & 80\% & 100\% \\
\hline
\multirow{2}{*}{Greedy, equal} & Coverage & $6.5\,{\scriptstyle\pm0.5}$ & $11.9\,{\scriptstyle\pm0.6}$ & $15.0\,{\scriptstyle\pm0.7}$ \\
\cline{2-5}
 & Over 0.7 & $6.9\,{\scriptstyle\pm1.0}$ & $13.1\,{\scriptstyle\pm0.8}$ & $16.5\,{\scriptstyle\pm0.9}$ \\
\hline
\multirow{2}{*}{Greedy, prop.} & Coverage & $16.1\,{\scriptstyle\pm0.0}$ & \dashuline{$13.8\,{\scriptstyle\pm0.8}$} & \dashuline{$18.1\,{\scriptstyle\pm0.8}$} \\
\cline{2-5}
 & Over 0.7 & \dashuline{$39.1\,{\scriptstyle\pm0.3}$} & \dashuline{$34.5\,{\scriptstyle\pm0.3}$} & \dashuline{$30.2\,{\scriptstyle\pm0.6}$} \\
\hline
\multirow{2}{*}{Max-min, equal} & Coverage & $4.4\,{\scriptstyle\pm0.4}$ & $7.6\,{\scriptstyle\pm0.2}$ & $7.8\,{\scriptstyle\pm0.2}$ \\
\cline{2-5}
 & Over 0.7 & $5.6\,{\scriptstyle\pm1.0}$ & $9.5\,{\scriptstyle\pm0.3}$ & $9.7\,{\scriptstyle\pm0.3}$ \\
\hline
\multirow{2}{*}{Max-min, prop.} & Coverage & \dashuline{$16.7\,{\scriptstyle\pm0.3}$} & $12.9\,{\scriptstyle\pm0.1}$ & $11.2\,{\scriptstyle\pm0.5}$ \\
\cline{2-5}
 & Over 0.7 & $37.7\,{\scriptstyle\pm0.8}$ & $30.8\,{\scriptstyle\pm0.7}$ & $28.1\,{\scriptstyle\pm0.7}$ \\
\hline
\end{tabular}
\end{table}

\begin{table*}[t]
\centering
\caption{Attack delta (\%) at 5\% budget, mean demand 30~Mbps, best cell count. Value $\pm$ 95\% CI. \textcolor{blue}{Blue}: most damaging configuration for that algorithm; \dashuline{dashed}: most damaged algorithm for that configuration. Higher is worse for users.}
\label{tab:heuristics-mean30}
\scriptsize
\setlength{\tabcolsep}{3pt}
\begin{tabular}{|l|l|cc|cc|cc|cc|}
\hline
\multirow{3}{*}{Scheduler / capacity sharing} & \multirow{3}{*}{Metric} & \multicolumn{4}{c|}{user-based ranking} & \multicolumn{4}{c|}{demand-based ranking} \\
\cline{3-6} \cline{7-10}
 &  & \multicolumn{2}{c|}{bottom} & \multicolumn{2}{c|}{top} & \multicolumn{2}{c|}{bottom} & \multicolumn{2}{c|}{top} \\
\cline{3-4} \cline{5-6} \cline{7-8} \cline{9-10}
 &  & uniform & waterfill & uniform & waterfill & uniform & waterfill & uniform & waterfill \\
\hline
\multirow{2}{*}{Greedy, equal} & Coverage & \textcolor{blue}{$11.9\,{\scriptstyle\pm0.6}$} & $11.2\,{\scriptstyle\pm0.7}$ & $1.7\,{\scriptstyle\pm0.0}$ & $1.9\,{\scriptstyle\pm0.4}$ & $11.8\,{\scriptstyle\pm0.7}$ & $11.8\,{\scriptstyle\pm0.7}$ & $1.7\,{\scriptstyle\pm0.0}$ & $6.5\,{\scriptstyle\pm0.4}$ \\
\cline{2-10}
 & Over 0.7 & $13.0\,{\scriptstyle\pm0.7}$ & $12.4\,{\scriptstyle\pm0.8}$ & $2.3\,{\scriptstyle\pm0.0}$ & $2.5\,{\scriptstyle\pm0.4}$ & \textcolor{blue}{$13.1\,{\scriptstyle\pm0.8}$} & \textcolor{blue}{$13.1\,{\scriptstyle\pm0.8}$} & $2.3\,{\scriptstyle\pm0.0}$ & $8.0\,{\scriptstyle\pm0.5}$ \\
\hline
\multirow{2}{*}{Greedy, prop.} & Coverage & \dashuline{\textcolor{blue}{$13.8\,{\scriptstyle\pm0.8}$}} & \dashuline{$12.9\,{\scriptstyle\pm0.9}$} & $12.6\,{\scriptstyle\pm0.0}$ & $8.9\,{\scriptstyle\pm0.0}$ & \dashuline{$13.7\,{\scriptstyle\pm0.9}$} & \dashuline{$13.7\,{\scriptstyle\pm0.9}$} & $12.6\,{\scriptstyle\pm0.0}$ & $12.3\,{\scriptstyle\pm0.0}$ \\
\cline{2-10}
 & Over 0.7 & \dashuline{$17.7\,{\scriptstyle\pm1.2}$} & \dashuline{$17.3\,{\scriptstyle\pm1.2}$} & \dashuline{$25.4\,{\scriptstyle\pm0.3}$} & $16.1\,{\scriptstyle\pm0.1}$ & \dashuline{$18.2\,{\scriptstyle\pm1.3}$} & \dashuline{$18.2\,{\scriptstyle\pm1.3}$} & \dashuline{$25.2\,{\scriptstyle\pm0.4}$} & \dashuline{\textcolor{blue}{$34.5\,{\scriptstyle\pm0.3}$}} \\
\hline
\multirow{2}{*}{Max-min, equal} & Coverage & \textcolor{blue}{$7.6\,{\scriptstyle\pm0.2}$} & $7.0\,{\scriptstyle\pm0.5}$ & $1.6\,{\scriptstyle\pm0.0}$ & $1.6\,{\scriptstyle\pm0.0}$ & \textcolor{blue}{$7.6\,{\scriptstyle\pm0.2}$} & \textcolor{blue}{$7.6\,{\scriptstyle\pm0.2}$} & $1.6\,{\scriptstyle\pm0.0}$ & $1.8\,{\scriptstyle\pm0.0}$ \\
\cline{2-10}
 & Over 0.7 & \textcolor{blue}{$9.5\,{\scriptstyle\pm0.3}$} & $8.7\,{\scriptstyle\pm0.7}$ & $2.2\,{\scriptstyle\pm0.0}$ & $2.4\,{\scriptstyle\pm0.0}$ & $9.4\,{\scriptstyle\pm0.3}$ & $9.4\,{\scriptstyle\pm0.3}$ & $2.2\,{\scriptstyle\pm0.0}$ & $2.5\,{\scriptstyle\pm0.0}$ \\
\hline
\multirow{2}{*}{Max-min, prop.} & Coverage & $10.1\,{\scriptstyle\pm0.3}$ & $9.4\,{\scriptstyle\pm0.7}$ & \dashuline{\textcolor{blue}{$12.9\,{\scriptstyle\pm0.1}$}} & \dashuline{$9.6\,{\scriptstyle\pm0.0}$} & $10.1\,{\scriptstyle\pm0.3}$ & $10.1\,{\scriptstyle\pm0.3}$ & \dashuline{\textcolor{blue}{$12.9\,{\scriptstyle\pm0.1}$}} & \dashuline{$12.7\,{\scriptstyle\pm0.1}$} \\
\cline{2-10}
 & Over 0.7 & $9.9\,{\scriptstyle\pm0.6}$ & $9.4\,{\scriptstyle\pm0.9}$ & $23.7\,{\scriptstyle\pm0.4}$ & \dashuline{$16.3\,{\scriptstyle\pm0.1}$} & $9.9\,{\scriptstyle\pm0.6}$ & $9.9\,{\scriptstyle\pm0.6}$ & $23.5\,{\scriptstyle\pm0.4}$ & \textcolor{blue}{$30.8\,{\scriptstyle\pm0.7}$} \\
\hline
\end{tabular}
\end{table*}

\subsection{Ablation Study}
\label{sec:ablation}

We evaluate three system choices and their impacts to get a more realistic overview of the numbers in the main result in Section~\ref{sec:results}. The choices are: The knowledge of the attacker about the demands, how loaded the system is when the attack lands (utilization), and the effect of satellite movement over time.

\subsubsection{Demand knowledge effect}
Both target selection and water-filling operate on a per-cell score $\kappa_i$, and the attacker's information determines which quantity that score can be. A white-box attacker who observes reported demand ranks by aggregate cell demand, $\kappa_i = D_i$. This is the quantity the system itself acts on: it sets the cell's beam target $q_i = \lceil D_i/W \rceil$, and under proportional sharing it also sets how the cell's delivered capacity is divided among its terminals. A black-box attacker sees only how many terminals are registered in each cell and must rank by population, $\kappa_i = U_i$, as a proxy. Population is a proxy for demand rather than a substitute for it: it correlates with $D_i$, so it usually orders cells similarly, but it enters the mechanism directly only under \emph{equal} sharing, where $U_i$ is the sharing denominator.

The distinction is observable in this deployment because per-terminal demand is heterogeneous; a cell with many light terminals and a cell with few heavy ones can occupy opposite ends of the two rankings. That gap has almost no effect when the budget is spread uniformly, since uniform spreading ignores cell levels entirely, but it becomes decisive once water-filling is used, because water-filling levels whichever score it is keyed on. Table~\ref{tab:heuristics-mean30} therefore reports both ranking bases side by side, and the two subsubsections below separate what the spread rule does from what the attacker needs to know to use it.

At the 5\% budget, detailed demand knowledge matters primarily for top targeting with water-filling under proportional sharing. Relative to population-based scoring, demand-based scoring increases the loss in users achieving at least 0.7 satisfaction from 16.1\% to 34.5\% for greedy scheduling and from 16.3\% to 30.8\% for max-min scheduling. Under equal sharing, the strongest bottom-targeting results are nearly unchanged: 13.0\% versus 13.1\% for greedy and 9.5\% versus 9.4\% for max-min. Thus, population counts are sufficient under equal sharing, whereas accurate demand information becomes important when proportional sharing and water-filling interact.

\subsubsection{Load Sensitivity}
Table~\ref{tab:utilization} shows that higher utilization does not uniformly increase the measured attack damage; instead, the response depends strongly on the sharing rule. Under equal sharing, damage increases with utilization for both schedulers. When spare capacity is available, the scheduler can absorb part of the adversarial demand by allocating additional beams. As the system approaches saturation, this headroom disappears, and injected users must compete directly with legitimate users for already limited capacity. The increase is steepest under greedy, whereas max-min begins to level off between 80\% and 100\% utilization. Proportional sharing exhibits a different pattern. The decrease in users attaining at least 0.7 satisfaction becomes smaller as utilization rises under both schedulers, while the coverage response is either decreasing or non-monotonic. This should not be interpreted as greater robustness at high load. Because the table reports damage relative to the no-attack baseline, a heavily loaded system already has fewer users above the relevant service thresholds, leaving less room for the attack to produce an additional measured decrease. At lower utilization, more users begin above these thresholds, and high-demand adversarial users can push a larger fraction below them.

\subsubsection{Time sensitivity}

\begin{figure}[t]
\centering
\begin{subfigure}[t]{0.48\columnwidth}
\centering
\includegraphics[width=\linewidth]{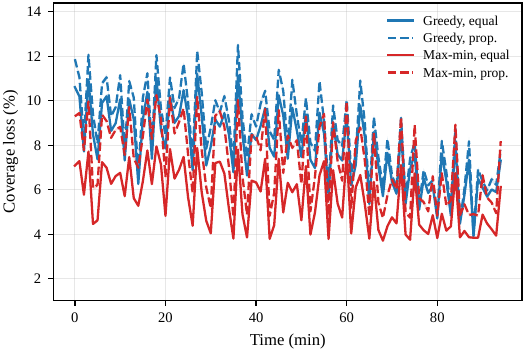}
\caption{Bottom heuristic.}
\label{fig:time-coverage-bottom}
\end{subfigure}\hfill
\begin{subfigure}[t]{0.48\columnwidth}
\centering
\includegraphics[width=\linewidth]{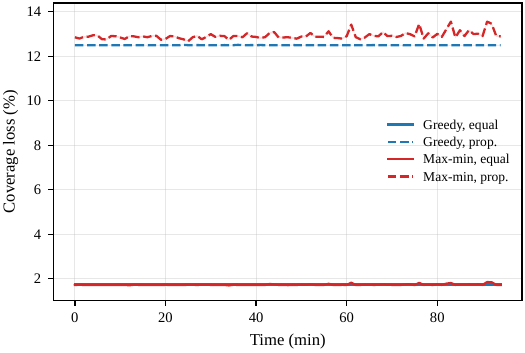}
\caption{Top heuristic.}
\label{fig:time-coverage-top}
\end{subfigure}
\caption{Effect of attack duration on coverage, at a representative 5\% adversary budget. Lines show the mean and shaded regions show uncertainty. The shape of the response depends on the scheduler, sharing rule, and targeting heuristic.}
\label{fig:time-coverage}
\end{figure}

Figure~\ref{fig:time-coverage} reveals a clear difference in temporal sensitivity between the two targeting heuristics. The coverage loss produced by top targeting remains nearly constant throughout the orbital cycle. Because this heuristic selects high-demand cells that remain high priorities for beam allocation, changes in satellite visibility rarely alter their relative importance or the attack's effectiveness. Bottom targeting is substantially more time-dependent: its coverage loss fluctuates sharply and generally changes as the constellation moves. These attacks target lightly loaded, low-priority cells whose allocations are more sensitive to small changes in available capacity and connectivity. As the number of satellites visible from the service area varies from 84 to 99 during the orbital cycle, changes in feasible beam assignments can determine whether these marginal cells receive service, producing the observed variation in attack damage. Thus, top targeting provides relatively stable damage across time, whereas the effectiveness of bottom targeting depends more strongly on the constellation state and should be evaluated over a complete orbital cycle rather than at a single snapshot.

\section{Conclusion}
\label{sec:conclusion}

This paper showed that denial-of-service attacks can disrupt the beam allocation in large LEO constellations. Both the sharing rule and scheduler affect vulnerability. Sharing has a strong effect at every budget: at a 5\% budget, proportional sharing causes up to a 34\% loss in users getting at least 0.7 satisfaction, compared with up to 13\% under equal sharing. The scheduler becomes more influential at higher budgets, and max-min makes the attack more costly by requiring it to span $1.4\times$--$2\times$ as many cells as greedy. However, max-min/equal also has the lowest no-attack coverage, showing that robustness must consider both relative degradation and absolute service.

From a threat-modeling perspective, no single heuristic is strongest in every scenario: lightly loaded cells are most vulnerable under equal sharing, heavily loaded cells are most vulnerable under proportional sharing, and effectiveness also varies with the system utilization factor and the satellite visibility. System designers should therefore combine attack-aware beam scheduling with robust intra-cell sharing instead of relying on routing defenses alone. Future work will develop adaptive attack heuristics that remain effective across allocation policies, traffic loads, and orbital states, and use them to design and evaluate practical defenses. Although our simulations omit admission control and attack detection, the results motivate practical defenses, including demand caps and monitoring sudden terminal activations. Future work will evaluate these mechanisms across additional constellations and traffic conditions to guide the design of more resilient LEO networks.

\bibliographystyle{ieeetr}
\bibliography{main}
\end{document}